\documentclass[12pt,a4paper]{article}
 \usepackage[dvips]{graphics}
 \usepackage{graphicx}
 \usepackage{afterpage}
 \usepackage{enumerate}
  \usepackage{longtable}

\newcount\longrefs
\def\aap{\ifnum\longrefs=1 {Astron.\ Astrophys.}\else
                           {A\hbox{\rm \&}A}\fi}
\def\aapr{\ifnum\longrefs=1 {Astron.\ Astrophys.\ Rev.}\else
                            {A\hbox{\rm \&}AR}\fi}
\def\aaps{\ifnum\longrefs=1 {Astron.\ Astrophys.\ Suppl.}\else
                            {A\hbox{\rm \&}A Suppl.}\fi}
\def\aj{\ifnum\longrefs=1 {Astron.\ J.}\else
                          {AJ}\fi}
\def\ao{\ifnum\longrefs=1 {Applied Optics}\else
                           {Appl.\ Opt.}\fi}
\def\aspcs{\ifnum\longrefs=1 {Astron.\ Soc.\ Pacific Conf. Series}\else
                           {ASP Conf.\ Ser.}\fi}
\def\apj{\ifnum\longrefs=1 {Astrophys.\ J.}\else
                           {ApJ}\fi}
\def\apjl{\ifnum\longrefs=1 {Astrophys.\ J. Lett.}\else
                            {ApJ}\fi}
\def\aplett{\ifnum\longrefs=1 {Astrophys.\ J. Lett.}\else
                            {ApJ}\fi}
\def\apjs{\ifnum\longrefs=1 {Astrophys.\ J. Suppl.}\else
                            {ApJS}\fi}
\def\apss{\ifnum\longrefs=1 {Astrophys.\ and Space Science}\else
                            {Astrophys.\ Space Sci.}\fi}
\def\araa{\ifnum\longrefs=1 {Ann.\ Rev.\ Astron.\ Astrophys.}\else
                            {ARA\hbox{\rm \&}A}\fi}
\def\azh{\ifnum\longrefs=1 {Astronomicheskii Zhurnal}\else
                            {Astron.\ Zhur.}\fi}
\def\baas{\ifnum\longrefs=1 {Bull.\ Am.\ Astron.\ Soc.}\else
                            {BAAS}\fi}
\def\bain{\ifnum\longrefs=1 {Bull.\ Astronom.\ Institutes Netherlands}\else
                            {Bull.\ Astr.\ Inst.\ Neth.}\fi}
\def\gca{\ifnum\longrefs=1 {Geochim.\ Cosmochim.\ Acta}\else
                           {Geochim.\ Cosmochim.\ Acta}\fi}
\def\grl{\ifnum\longrefs=1 {Geophys.\ Res.\ Lett.}\else
                           {Geoph.\ Res.\ Lett.}\fi}
\def\iaucirc{\ifnum\longrefs=1 {IAU Circulars}\else
                          {IAU Circ.}\fi}
\def\ip{\ifnum\longrefs=1 {in press}\else
                          {in press}\fi}
\def\jgr{\ifnum\longrefs=1 {J.\ Geophys.\ Res.}\else
                           {J.\ Geophys.\ Res.}\fi}
\def\jrasc{\ifnum\longrefs=1 {J.\ Royal Astron.\ Soc.\ Canada}\else
                           {JRAS Can.}\fi}
\def\mnras{\ifnum\longrefs=1 {Mon.\ Not.\ Roy.\ Astron.\ Soc.}\else
                             {MNRAS}\fi}
\def\nat{\ifnum\longrefs=1 {Nature}\else
                           {Nat}\fi}
\def\pasj{\ifnum\longrefs=1 {Pub.\ Astron.\ Soc.\ Japan}\else
                            {PASJ}\fi}
\def\pasp{\ifnum\longrefs=1 {Pub.\ Astron.\ Soc.\ Pacific}\else
                            {PASP}\fi}
\def\physscr{\ifnum\longrefs=1 {Physica Scripta}\else
                            {Phys.\ Scrip.}\fi}
\def\planss{\ifnum\longrefs=1 {Planetary \& Space Science}\else
                            {Plan. \& Space Sci.}\fi}
\def\procspie{\ifnum\longrefs=1 {Proc.\ SPIE}\else
                            {Proc.\ SPIE}\fi}
\def\qjras{\ifnum\longrefs=1 {Quarterly J.\ Royal Astron.\ Soc.}\else
                            {QJRAS}\fi}
\def\sa{\ifnum\longrefs=1 {Soviet Astron..}\else
                               {Sov.\ Astron.}\fi}
\def\skytel{\ifnum\longrefs=1 {Sky \& Telescope}\else
                            {Sky \& Tel.}\fi}
\def\solphys{\ifnum\longrefs=1 {Solar Phys.}\else
                               {Sol.\ Phys.}\fi}
\def\ssr{\ifnum\longrefs=1 {Space Science Rev.}\else
                               {Space\ Sci.\ Rev.}\fi}

\begin{document}

 \title{\bf Formation depths  of Fraunhofer lines
}

 \author{\bf E.A. Gurtovenko, V.A. Sheminova}
 \date{}

 \maketitle
 \thanks{}
\begin{center}
{Main Astronomical Observatory, National Academy of Sciences of
Ukraine
\\ Zabolotnoho 27, 03689 Kyiv, Ukraine\\ E-mail: shem@mao.kiev.ua}
\end{center}

 \begin{abstract}
We have summed up our investigations performed in 1970--1993. The
main task of this paper is clearly to show processes of formation
of spectral lines as well as their distinction by validity and by
location. For 503 photospheric lines of various chemical elements
in the wavelength range 300--1000 nm we list in Table the average
formation depths of the line depression and the line emission for
the line centre and on the half-width of the line, the average
formation depths of the continuum emission as well as the
effective widths of the layer of the line depression formation.
Dependence of average depths of line depression formation on
excitation potential, equivalent widths, and central line depth
are demonstrated by iron lines.

\end{abstract}

\section{Historic aspect of the problem}
     \label{S-Introduction}

In the 60 years the quantitative studies of the solar atmosphere
demanded knowledge of its physical characteristics at different
depths. The  majority of these characteristics were derived from
the Fraunhofer lines observed. Naturally it was assumed that the
atmospheric parameter derived from the specific Fraunhofer line
must be referred to the formation depth of the line. Therefore,
the question arose about the average depth of  formation of
spectral lines.

Recall, that the Fraunhofer line or  spectral absorption line is a
weakening of the intensity of continuous spectrum of radiation,
resulting from deficiency of photons in a narrow frequency range,
compared with the nearby frequencies. This deficiency is created
by some particles that absorb the photons in the narrow frequency
range (i.e. by selective absorption). The weakening of the
emergent radiation at the line frequency is often called the line
depression. At the solar surface for the otical depth
$\tau=0$  the emergent line depression is defined by:
\begin{eqnarray}
 \label{Eq:0}
 D_l(0) =I_c(0)-I_l(0),
\end{eqnarray}
where $I_l(0)$ is the intensity of the  emergent line emission,
$I_c(0)$ is the intensity of the  continuous emission  at the
frequency considered  in the case if this line would be absent.

In the 60 years in Kiev  we also began to study  of the structure
and dynamics of the solar atmosphere. We believed that the line
formation depth  is determined by the contribution function (CF)
to the emergent line emission $I_l$. It is easily calculated.
The solution of the transfer equation for the optical depth
$\tau=0$ gives an expression for the emergent  intensity  at  the
line frequency:
\begin{eqnarray}
 \label{Eq:1}
 I_l(0)=\int\limits_{0}^\infty S \exp(-\tau/\mu)d\tau/\mu,
\end{eqnarray}
where $\tau=\tau_l+\tau_c$ is the total (line + continuum) optical
depth. Here and further in the text the subscript $l$ means the
quantity refers to the line and $c$  to the continuum. $S$ is the
total source function, and $\mu=\cos \theta$ is the cosine of the
angle between the direction of the radiation and the normal to the
surface.  For simplicity we have adopted $\mu = 1$ in  the
subsequent formulas. Taking into account the following
relationships
\begin{eqnarray}
 \label{Eq:4}
S=\frac{S_c + \eta S_l}{1+\eta},~~ \eta=\frac{\kappa_l}{\kappa_c},~~d\tau= (1+\eta){d\tau_c},
\end{eqnarray}
one can write the equation (\ref{Eq:1}) in the scale of integration $\tau_c$:
\begin{eqnarray}
 \label{Eq:3}
 I_l(0)=\int\limits_{0}^\infty (S_c+\eta S_l)\exp(-(\tau_l+\tau_c))d\tau_c
 = \int\limits_{0}^\infty F_Ed\tau_c.
 \end{eqnarray}
Here, ${\kappa_c},~{\kappa_l}$ are the  coefficients of the
continuous and selective absorption, $F_E$ is  the contribution
function to emergent line emission or {\it the emission CF}. It
presents  the contributions  from atmospheric layers located
between $\tau_c$ and $\tau_c+ d\tau_c$ to the  intensity of the
emission observed in the line frequency. The emergent intensity of
the  nearby continuum  at the frequency considered is defined by
following formula:
 \begin{eqnarray}
 \label{Eq:3a}
 I_c(0)=\int\limits_{0}^\infty S_c\exp(-\tau_c)d\tau_c
 = \int\limits_{0}^\infty F_C d\tau_c,
 \end{eqnarray}
where $F_C$ is the  contribution function to emergent continuous
emission.

The application of the emission CF  for calculation of the average
depth  of  formation of the absorption line in our practice  has
convinced us that this is wrong. The average depths obtained are
practically identical for all very weak lines with different
excitation potentials at the similar wavelength. At that time it
was well known that the very weak lines with higher excitation
potentials are formed in the deeper layers of the photosphere
compared with the similar lines with lower excitation potentials.
We were forced to find  a new method  for determining the average
depth  of formation of spectral line.

In the 70  years, we began a detailed study of this problem
together with well-known astrophysicist C. de Jager
\cite{1974SoPh...37...43G}. We  showed that  it is important to
distinguish the formation regions  of the line emission and line
depression. They may be different. Also we showed that it should
use the depression contribution function based on the Unsold's
weighting function  to determine the average depth of formation of
the line depression.  We used  the weighting  function under LTE
conditions which was written as:
\begin{eqnarray}
 \label{Eq:7b}
g^{\prime}(\tau_c)=\int\limits_{\tau_c}^\infty B \exp(-\tau_c)d\tau_c -B\exp(-\tau_c).
\end{eqnarray}
Here, $g^{\prime}$ is differently defined than the
customary weighting  function $g$ derived by Unsold
\cite{1932ZA......4..339U}. By multiplying  the Unsold's weighting
function $g$ with $I_c(0)$, we has obtained $g^{\prime}$. With the
function $g^{\prime}$ the expression for the emergent line
depression in the intensity units has the following form:
\begin{eqnarray}
 \label{Eq:6}
D_l(0)=\int\limits_{0}^\infty g^{\prime} \eta\exp(-\tau_l)d\tau_c =\int\limits_{0}^\infty F_D d\tau_c.
\end{eqnarray}
Here, $F_D$ is {\it the depression CF}. If  the absorbing
particles are absent ($\kappa_l=0,~ \eta=0, $ and $F_D=0$), the
contributions to the line depression equal zero. One can  divide
(\ref{Eq:6}) by  $I_c(0)$  to  obtain   the expression for the
relative depression ($R_l(0)=D_l(0)/I_c(0)$)  and the contribution
function to the relative depression ($F_R=F_D/I_c(0))$.  The
function $F_R$ corresponds to the contribution function derived by
Unsold \cite{1932ZA......4..339U} and  Pecker
\cite{1951AnAp...14..115P}.

In our next paper \cite{1974AZh....51.1032G} we   calculated the
average depths of formation for many weak spectral lines using the
function $F_D$. The obtained depths of formation of weak lines
showed a clear dependence on the excitation potential.  Thus, we
confirmed the validity of the determination of the average
formation depth with the contribution function $F_D$ as well as
$F_R$.

After the our papers
\cite{1974AZh....51.1032G,1974SoPh...37...43G}, a lot of  other
papers devoted to this problem have appeared. The opinions of the
authors of the papers were strongly divided. Some authors (e.g.,
Demidov \cite{1980IGAFS..52....3D,1983IGAFS..65...37D}) supported
our concept of the depression CF. Other authors (Babiy and
Rykalyuk \cite{1981AZh....58..825B}, Buslavsky
\cite{1969IzKry..39..317B}, Lyubimkov \cite{1976IzKry..55..164L},
and etc.) defended the concept of the emission CF from more great
tenacity. Third authors (Makita \cite{1977SoPh...51...43M},
Beckers and Milkey \cite{1975SoPh...43..289B}, and etc.) have
proposed new methods of calculating the formation depth of
absorption line. There were objective reasons to explain why our
concept was not immediately accepted and the problem  of the
formation depths was discussed for a long time.  We can call the main
reasons.

1. The inertia of thinking. Having solved the problem of radiative
transfer in the atmosphere of the Sun and stars, classical
astrophysics determined the line formation depth by the simple
way. The first definition  was based on the emergent line
emission. Minnaert \cite{1948BAN....10..399M} assumed  the average
depth of formation of absorption line is equal to the depth which
divides the $F_E$-curve in to equal halves. The second definition
was based on the approximation that  the emergent line intensity
is equal to the source function  at the average line formation
depth ${<}\tau_c{>}$. Under the LTE conditions this is
\begin{eqnarray}
  \label{Eq:6a}
I_l(0)= B({<}\tau_c{>}).
\end{eqnarray}
Using a photospheric model or data on the centre-to-limb variation
of the absolute continuum $I_c$, one can   easily obtain
${<}\tau_{c}{>}$. If the Plank function $B$ varies linearly with
$\tau_c$ one can use the important Eddington-Barbier approximation
$I_l(0)=B(\tau_c=\mu)$. Since then, many astrophysics
automatically use the solution of the transfer equation for the
emergent line intensity $I_l$ to determine the average depth of
formation of spectral line with the help of the emission CF
(although it was fundamentally wrong).

2. In the 70s and 80s the astrophysics did not take into account
the results of earlier works on this issue. In the 50--60s
Minnaert \cite{1948BAN....10..399M} and de Jager \cite{Jager52}
did not regard the emission CF for  determination of  the average
depth of formation of the Fraunhofer line. While Elste
\cite{1955ZA.....37..184E} and Ruhm \cite{1969A&A.....3..277R}
have pointed to the need to use the Unsold-Pecker CF.

3. The Unsold-Pecker CF  was  derived by mathematical
transformation to simplify the description of the Fraunhofer
lines.  Unsold \cite{1932ZA......4..339U} and later Pecker
\cite{1951AnAp...14..115P} did  not explain the physical meaning
of the function $F_R$. Apparently, in our first paper
\cite{1974SoPh...37...43G} the physical sense of the function
$F_D$ was not convincingly substantiated. It was masked by the
fact that we used the LTE condition ($S_c=S_l=B(T)$). If LTE is
absent
 the weighting function
 \begin{eqnarray}
 \label{Eq:7}
g^{\prime}(\tau_c)=\int\limits_{\tau_c}^\infty S_c \exp(-\tau_c)d\tau_c -S_l\exp(-\tau_c).
\end{eqnarray}
At that rate it is evident that the left term of (\ref{Eq:7}) the
multiplied by $\eta \exp(-\tau_l)$ presents the contribution to
the absorbtion by selectively absorbing particles at the depth
$\tau_c$, while  the right term (with negative sign) multiplied by
$\eta \exp(-\tau_l)$  presents the contribution  to the
re-emission by selectively absorbing particles.  Such explanation
was absent in our first article \cite{1974SoPh...37...43G}, and
this is also not facilitate the timely approval of our concept.

Due to the current confusing situation, we published  a series of
new papers \cite{1978SoPh...58..241G}--\cite{1991SoPh..136..239G}
and \cite{1992KFNT....8...44S}--\cite{1978AAfz...36...32S}. It
should note, at the same time  Sarychev \cite{1986SvA....30..329S}
supported the concept  of  the depression CF and  actively
cooperated with us.  Magain \cite{1986A&A...163..135M}
presented the correct solution of the transfer equation for the
relative line depression  $R_l(0)$ and obtained  the
rigorous expression for the CF to the relative line depression.
Gurtovenko and Sarychev \cite{1988AZh....65..653G} have shown that
the depression CF obtained by Magain is identical to  the
Unzold-Pecker CF.

By the end of 80 years, we were convinced  that the problem of the
formation depths of spectral lines  has already been solved.
Nevertheless, a series of new papers have appeared.  Achmad
\cite{1992SoPh..138..411A} raise the issue again and proposed a
new method which confirms the correctness of our concept.
Grossmann-Doerth  et al. \cite{1988A&A...204..266G}, Ruiz Cobo and
Toro Inesta \cite{1994A&A...283..129R}, and Staude
\cite{1996SoPh..164..183S} developed the concept of depression
functions for four Stokes profiles of the absorption lines.

The purpose of this paper to summarize the results of our
researches on this issue, to demonstrate the features of the
different  contribution functions for different lines and to show
the localization of the absorption and emission processes
involved in the formation of the lines. Furthermore,  we want to
demonstrate  the average depth of  formation of different lines in
solar spectrum. The  lines selected  by us are often used in the
interpretation of many phenomena in the solar atmosphere,
therefore  the knowledge of the average depths of formation of
emission and depression of these lines can be useful for
astrophysics.


\section{Formation of  absorption line}
     \label{ processes}

\begin{figure}
   \centering
   \includegraphics[width=9cm]{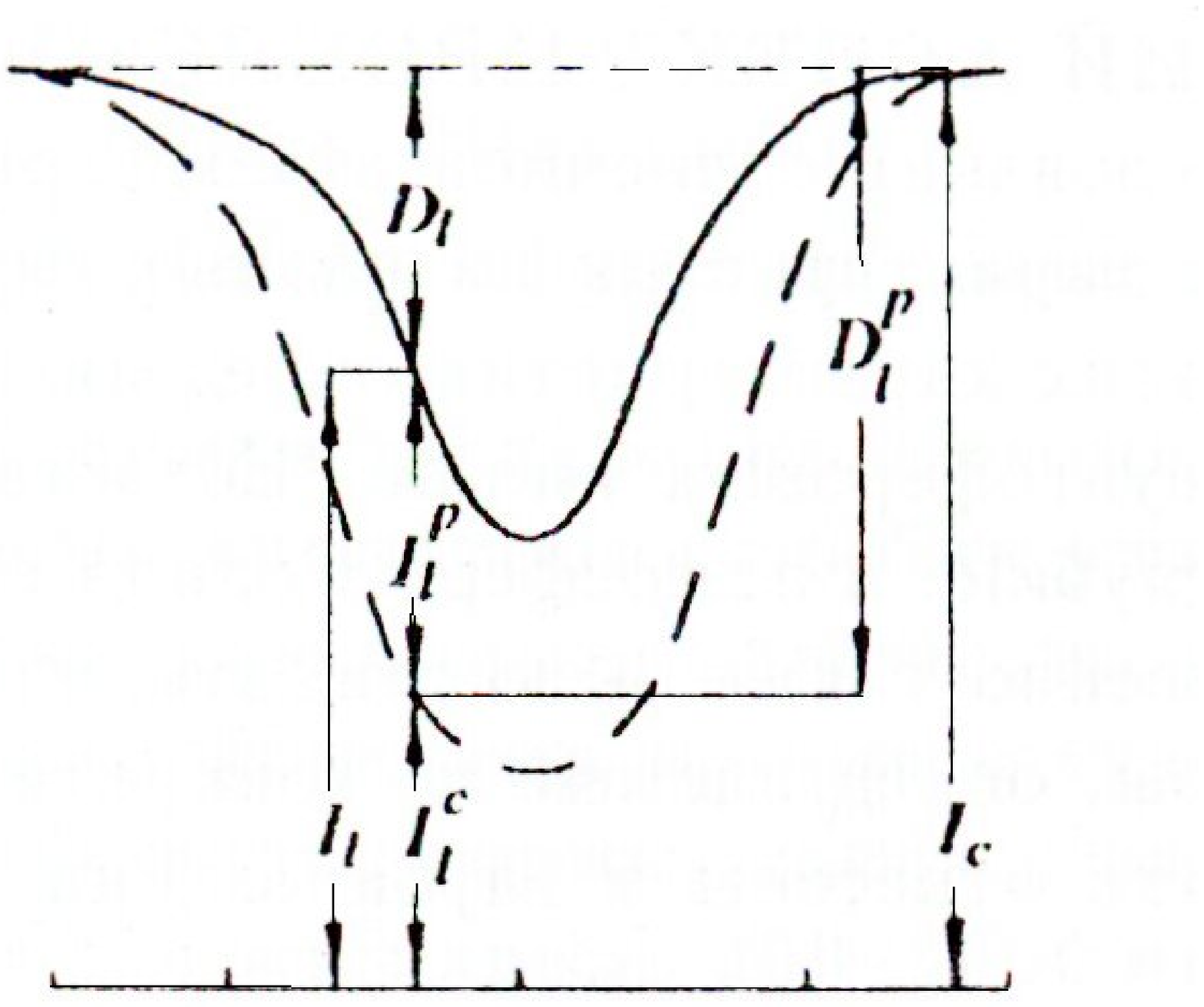}
   \caption
{The quantities of the contributions to observed  profile of a
Fraunhofer line (solid line) connected with the  emission and
depression processes forming the line and continuum. The intensity
of observed line emission $I_l = I_l^c+I_l^p$. The observed
profile of emergent line depression $D_l=D_l^p-I_l^p$. While $I_c$
is the intensity of the nearby continuum.  If the selective
absorbing particles did not produce  the proper emission
($I_l^p=0$), the observed line would have the profile shown by a
dashed line   }
    \label {F:1}
    \end{figure}

The physical sense of the depression contribution function becomes
perfectly clear, when we examine the specific processes forming a
absorption line. It is well known that  absorption line originates
as a consequence of a chain of absorption and emission processes
caused by  the presence selectively  absorbing particles in the
photospheric layers. Fig.~\ref{F:1} demonstrates a schematic ratio
of the quantities that characterizes the processes forming the
Fraunhofer line. Let's look these processes separately from each
other and present them quantitatively by the physical approach.

1. {\bf Selective absorption}. If we assume that selectively absorbing
particles  only absorb,  the processes of selective re-radiation
are absent, i.e., $\epsilon_l=0$ and $S_l=0$, then the portion of
radiation absorbed selectively at the depth $z$  within
an elementary  layer $\Delta z $  at  the line frequency will
equal to
\begin{eqnarray}
\label{Eq:8}
 \Delta D_l^p(z) = I_c(z) \exp(-\tau_l(z))\exp(-\tau_c(z))\kappa_l(z)\Delta z.
 \end{eqnarray}
Here $\exp(-\tau_l(z)) \exp(-\tau_c(z))$ is the factor of the
radiation attenuation  by the selective absorbing particles and
continuous absorption. After substitution of the expression for
the intensity of continuous radiation coming from the level
$\tau_c$
\begin{eqnarray}
\label{Eq:9}
I_c(\tau_c)=\exp(\tau_c) \int\limits_{\tau_c}^\infty S_c(\tau_c) \exp(-\tau_c) d \tau_c,
\end{eqnarray}
and  integration in the $\tau_c$ scale, we obtain the expression
for  amount of the radiation absorbed selectively  for $\tau_c=0$:
\begin{eqnarray}
\label{Eq:10}
 D_l^p(0)=\int\limits_{0}^\infty \left[  \int\limits_{\tau_c}^\infty S_c(\tau_c) \exp(-\tau_c)d\tau_c \right] \exp(-\tau_l)\eta d \tau_c.
\end{eqnarray}
We  call  the quantity $ D_l^p$  as {\it the proper line
depression}.

2. {\bf Selective re-emission}. Under real conditions
$\epsilon_l\neq0$ and $S_l \neq 0$. Some portion of the proper
line depression is be re-radiated in the line frequency. We
call it as {\it the  proper line emission}.   Analogously to
(\ref{Eq:8}),   the intensity of   proper line emission  equals:
\begin{eqnarray}
\label{Eq:11}
\
\Delta I_l^p(z)=\epsilon_l(z) \exp(-\tau_l(z)) \exp(-\tau_c(z))\Delta z.
\end{eqnarray}
Taking into account $\epsilon_l = S_l\kappa_l= S_l\kappa_c\eta $
and integrating over $\tau_c$, we obtain  the analytical
expression  for the emergent  intensity of   proper line emission
at the surface for  the  frequency within a spectral line:
\begin{eqnarray}
\label{Eq:12}
 I_l^p(0)=\int\limits_{0}^\infty  S_l \exp(-(\tau_c+\tau_l)) \eta d\tau_c.
\end{eqnarray}

3. {\bf  Continuous emission at the frequency within the spectral
line}. The selectively absorbing particles are not completely
"eat"  the continuous radiation  coming out in the line frequency.
Some part of the radiation comes out. We call it as  {\it the
continuous emission passed through the line}.  The analytical
expression  for the intensity of the continuous emission passed
through the line can be obtained  by the same simple way:
 \begin{eqnarray}
 \label{Eq:13}
 \Delta I_l^c(z)=\epsilon_c (z) \exp(-\tau_l(z))\exp(-\tau_c(z)) \Delta z.
\end{eqnarray}
After substitution of  $\epsilon_c =S_c\kappa_c$ and integration
over $\tau_c$, we obtain:
 \begin{eqnarray}
 \label{Eq:14}
 I_l^c(0)=\int\limits_{0}^\infty  S_c \exp(-(\tau_c+\tau_l)) d\tau_c.
\end{eqnarray}

4. {\bf Emergent line emission}  is the sum of the proper line
emission   and  the continuous emission passed through the line:
\begin{eqnarray}
 \label{Eq:15}
 I_l(0)= I_l^p(0)+I_l^c(0)= \int\limits_{0}^\infty  (S_c+ \eta S_l  ) \exp(-(\tau_c + \tau_l)) d \tau_c.
\end{eqnarray}
The integrand in  (\ref{Eq:15}) is the emission contribution
function $F_E$ (\ref{Eq:3})  derived from the solution of the
transfer equation for the emergent line intensity.

5. {\bf Observed line depression} is also formed by two processes.
The first process forms the proper line depression   and the
second  forms the proper line emission.  The   physical sum (or
the mathematical difference) of  the intensities  of the proper
depression and emission at the line frequency gives  the
expression for the  {\it  observed line depression}:
\begin{eqnarray}
\label{Eq:16}
 D_l(0)=D_c^p (0)- I_l^p(0) = \int\limits_{0}^\infty \left [ \int\limits_{\tau_c}^\infty S_c \exp(-\tau_c)d\tau_c - S_l   \exp(-\tau_c) \right]\eta \exp(-\tau_l) d \tau_c.
\end{eqnarray}
The integrand in  (\ref{Eq:16}) is the depression CF. Under  the
LTE approximation it corresponds to the expression
$F_D$ (\ref{Eq:6})  derived with the help of  the weighting functions
$g^\prime$.
  \begin{figure}
    \centering
   \includegraphics[width=10cm]{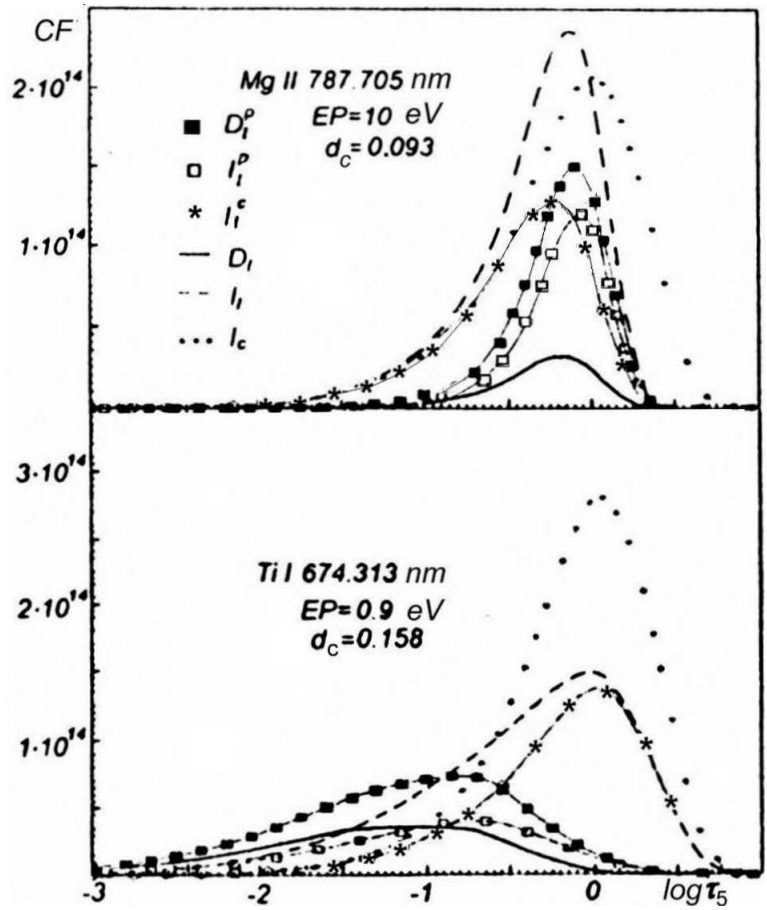}
   \caption
{Contribution functions  to the quantities $D_l^p, ~I_l^p,~
I_l^c,~ D_l, I_l,~ I_c$  at the center of  weak Fraunhofer  lines
}
  \label {F:2}
  \end{figure}

  \begin{figure}
    \centering
   \includegraphics[width=10cm]{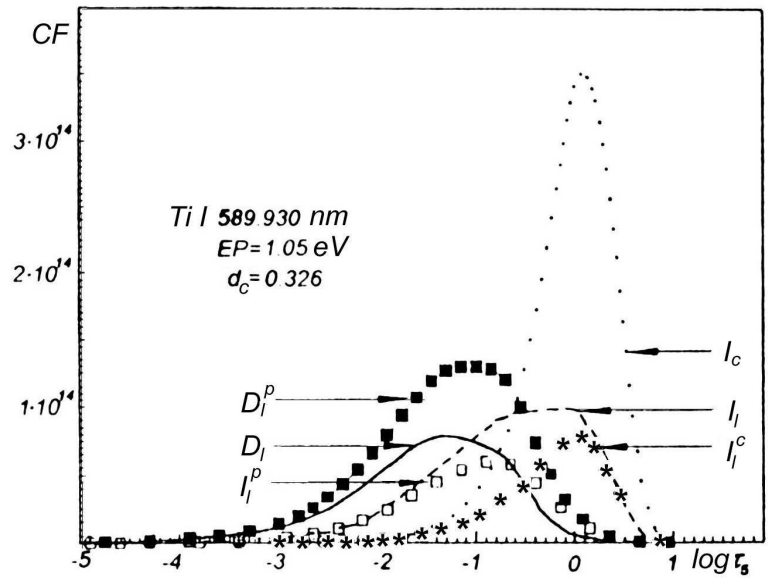}
   \caption
{The contribution functions at the center of  a moderate  Fraunhofer  line}
  \label {F:3}
  \end{figure}

\section{Features of the contribution functions }

In the spectral analysis one can use  the emission and depression
CFs as well as  the CFs to proper line emission and proper line
depression. All the CFs  carry the valuable information.   The
information on an amount and location of the proper line emission,
proper line depression, or  the continuous emission passed by the
line has  great practical significance in the study of the
relationship between the observed phenomena,  parameters of the
solar atmosphere, and spectral lines.   For example, the
brightness  contrasts  observed in different lines are compared
with such phenomena as concentration of magnetic field.  To study
such phenomena we must use the CF to the proper line emission
$I_l^p$, while  to study the photospheric velocity fields derived
from the Doppler lineshifts we should use the depression CF to
$D_l$.

\begin{figure}
    \centering
   \includegraphics[width=10cm]{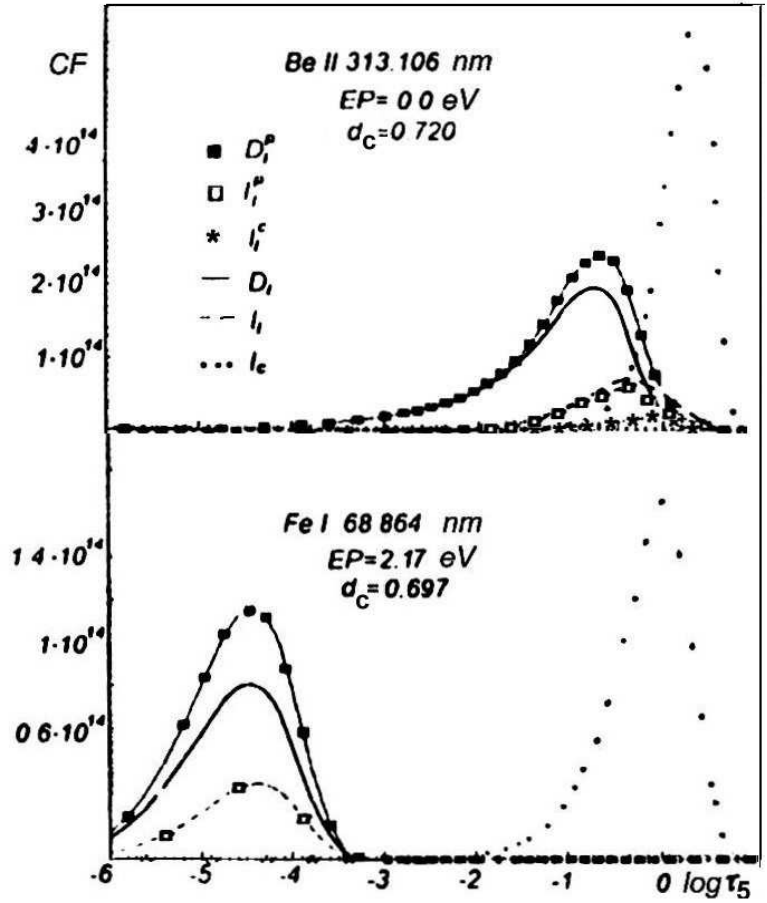}
   \caption
{The contribution functions at the center of strong  Fraunhofer  lines}
  \label {F:4}
  \end{figure}

We calculated the CFs under LTE conditions with photospheric model
HOLMU for the center of  different  lines at the solar disc center
in logarithmic scale of optical depths $\log \tau_5$. The scale $
\log\tau_5$ is more convenient to analyze  spectral lines in
stellar atmospheres. Figs~\ref{F:2}--\ref{F:4} are shown the CFs
calculated. Specific CF at specific $\log \tau_5$ shows   the
contribution of the layer situated at this depth  to the quantity
by the  specific  process.

The emergent intensity of continuum ($I_c$) is effectively formed
in the deepest layers of the photosphere. The formation region of
the intensity of continuous emission passed through the line
($I_l^c$)  is always located within the region of formation of the
nearby continuum. While  the formation regions of the proper line
depression ($D_l^p$) the proper line emission ($I_l ^p$) as well
as the  line depression ($D_l$) are not associated with the
formation of the continuum. They are always  associated with the
location of selectively absorbing particles. Therefore the
effective contributions to $D_l^p$, $I_l ^p$, and $D_l$, are
formed  in the same layers slightly offset from each other.
The CF to the emergent line intensity, $I_l$, may in principle
have two peaks,  since it  includes the contributions  associated
with selective re-emission  and  the continuum emission passed
through the line. Fig.~\ref{F:3} demonstrates the contributions to
$I_l^p$ and $I_l^c$ that are formed in different layers in the
case of a moderate line, while in the case of the very weak line
(Fig.~\ref{F:2}, Mg II 787.705~nm) they are formed practically in
the same layers. The magnitude and  location  of the maximum  of
the CFs  basically depends on the line strength, but they also
depend on  other line parameters. Let's  consider the CFs
calculated   for the weak, moderate, and strong lines,
separately.

{\it Weak lines}. Figure~\ref{F:2} shows the CFs calculated  for
very weak line of Mg II (787.705~nm). Due to small optical
thickness in very weak line its CF   clearly indicates the
location of the region of effective formation of the line. Another
important feature is that  the very weak line  is formed very
deeply in the photosphere. The greater the excitation potential,
the deeper and narrower the line formation region. Important
feature of the very weak lines with very high EP is   a large
amount of  the proper line emission  $I_l^p$  and the proper line
depression $D_l^p$ compared with  the observed line depression
$D_l$. Large amount of the proper line emission is remarkable
feature to explore of the solar and stellar atmospheres.

The weak lines with low excitation potentials also draw attention
to themselves. For example, the Ti I 674.313 line (Fig.~\ref{F:2})
has the  proper line emission  $I_l^p$ is comparable to the
observed line depression $D_l$. It is interesting that for this
line the location of the line emission differs significantly from
the formation region  of the line  depression. For weak lines with
high excitation potentials the difference is much smaller. The
formation region of resonance weak lines is located slightly
higher than usual weak lines with low excitation potentials, while
weak lines of molecules are formed within the highest layers of
the photosphere \cite{1986SvA....30..329S}.

{\it Moderate lines}. As we can see in Fig.~\ref{F:3}, the
formation regions  of  the proper line emission, the proper line
depression, and observed line depression  in the case of moderate
lines are shifted to higher layers of the photosphere as compared
with weak  lines  with the same excitation potentials and
wavelengths. A known fact of the decrease of the depth of
formation region of the moderate lines in comparison with the weak
lines is confirmed. For moderate line the amount of  the continuum
emission passed through the line is clearly reduced. The
dependence of  formation depth of moderate line on the excitation
potential is the same as that for weak lines.

{\it Strong lines}. It is known, that there is no any strong lines
with high excitation potentials in the solar spectrum, which could
be formed in the deep layers of the photosphere. Figure~\ref{F:4}
shows the CFs calculated  for strong  lines with low excitation
potentials. We  found  only one resonance line (313.106 Be II),
which is formed in the deep layers. Partly, it is a consequence of
its short wavelength. This line  passes a small portion of the
continuum emission, whereas the other strong lines  practically do
not transmit continuous radiation outwards. Therefore, the
emergent  line intensity  $I_l$ of  strong lines usually equals
the intensity of the proper line emission  $ I_l^p$. Its amount is
significantly  smaller compared with the observed line depression
$D_l$ or  the proper line depression  $D_l^p$. Thus, the small
value of the observed central intensity $I_l$  in the strong lines
is determined only by its proper emission re-emitted  by the
selectively absorbing particles. Hence it becomes clear why the
average depths of formation of emergent line emission and line
depression are the same in the strong lines.  Fig.~\ref{F:4} also
shows the CFs for strong atypical chromospheric line (868.864 nm
Fe I), which is formed in the uppermost layers of the photosphere
or practically in the low chromosphere. The location of the region
of effective formation of the strong line depends mainly on the
line strength.

We did not consider the strongest line, since their central part
is formed in the absence of the LTE conditions. Under non-LTE, the
relation between the population of the upper and lower levels
deviates dramatically from LTE equilibrium and can be quite
different at different depth in the photosphere. The shape and
magnitude of the CFs to $D_l^p,~I_l^p,~ D_l$  can significantly
change. In the case if the proper line emission $I_l^p$ is very
strong and the line depression  $D_l$ is negative, then the
emission  appears at the line center. Such phenomenon can occur in
the spectrum of the solar flares as the appearance of the emission
lines on the background of the continuum.

\section {The average depth of formation of a spectral line}

Despite the fact that nowadays every researcher has a computer and
one can calculate the average depth of formation of any line, and
despite the fact that Gurtovenko and Kostyk
\cite{1989KiIND.........G} have published the mean geometrical
depths  of  formation of the line center  for 1958 Fraunhofer
lines of different elements, we decided to present in accessible
form more complete information about the average depths of the 503
reference spectral lines. These lines can serve as probes to
diagnose the physical conditions of the various layers in the
photosphere.

Since the depression contribution function $F_D$ (\ref{Eq:6})
determines the weight of the relevant layers in contributing to
the emergent line depression, one may define a average  depth of
formation of the line  at a given wavelength profile point
($\lambda$)   at  a  position $\mu$ on the solar (stellar) disk by
the following formula:
 \begin{eqnarray}
\label{Eq:17}
{<} x{>}_{\mu}~= \int\limits_{-\infty }^\infty  x {F_D}(x,\mu) d x~ { /}\int\limits_{-\infty }^\infty  {F_D}(x,\mu) d x.
\end{eqnarray}
Here, $x$ can be equal to  $\tau_5$, $\log \tau_5$ or $z$.
The effective depth of formation of  the entire line profile at
the  position $\mu$  is determined by averaging over the whole
profile:
\begin{eqnarray}
\label{Eq:17b}
{<} x{>}_{\rm \mu,~eff}~= \int\limits_{\lambda_1 }^{\lambda_2}{<} x{>}_{\mu} R (\lambda,\mu)d \lambda~ { /}\int\limits_{\lambda_1 }^{\lambda_2} R (\lambda,\mu)d \lambda,
\end{eqnarray}
where $R (\lambda,\mu)$ is  the relative line depression at the
profile point $\lambda$ and at the position $\mu$, and
$\lambda_1,~\lambda_2$ are initial and final wavelength points of
the line profile.

Averaging over the solar disk one can obtain the mean depth of
formation of the line depression at a given profile point
$\lambda$ for the Sun as a star (or for any star):
\begin{eqnarray}
\label{Eq:17a}
{<} x{>}^*~= \int\limits_{0 }^{1}{<} x{>}_{\mu}\mu d \mu,
\end{eqnarray}
and the effective formation depth of  the entire line profile for
the Sun as a star:
\begin{eqnarray}
\label{Eq:17c}
{<} x{>}^{*}_{\rm eff}~= \int\limits_{\lambda_1 }^{\lambda_2}  {<} x{>}^{*} R^{*} (\lambda)d \lambda~ { /}\int\limits_{\lambda_1 }^{\lambda_2} R^{*} (\lambda,\mu)d \lambda~,
\end{eqnarray}
where  the superscript $*$  means the quantity averaged over the
solar (stellar) disk.

Analogously, one can  compute  the mean depth for the emergent
line emission using the emission contribution function  $F_E$
(\ref{Eq:3}) in (\ref{Eq:17}--\ref{Eq:17c}). The mean depth of
formation the nearby continuum  for each line one can  compute
using the CF to continuum  $F_C$ (\ref{Eq:3a}).  The CFs to line
depression, line emission, and continuum do not change sing,
therefore the  formulas (\ref{Eq:17}--\ref{Eq:17c}) give reliable
results.

The effective width of the region of formation of line depression
at the line center and the center of the solar disk ($\mu=1$)  one
can define as:
\begin{eqnarray}
\label{Eq:18}
\Delta x_{\rm eff}= \int\limits_{-\infty }^\infty (x-{<}x{>})^2 F_D(x)d x ~ { /}\int\limits_{-\infty }^\infty  F_D(x) d x,
\end{eqnarray}
where  ${<}x{>}$ is the average  depth of formation of the line
depression  at the line center. Thus, $\Delta x_{\rm eff}$ is the
width  of the layer in which occurs  the main part of all
contributions to  the observed line depression at the line center.

\begin{figure}
    \centering
   \includegraphics[width=10cm]{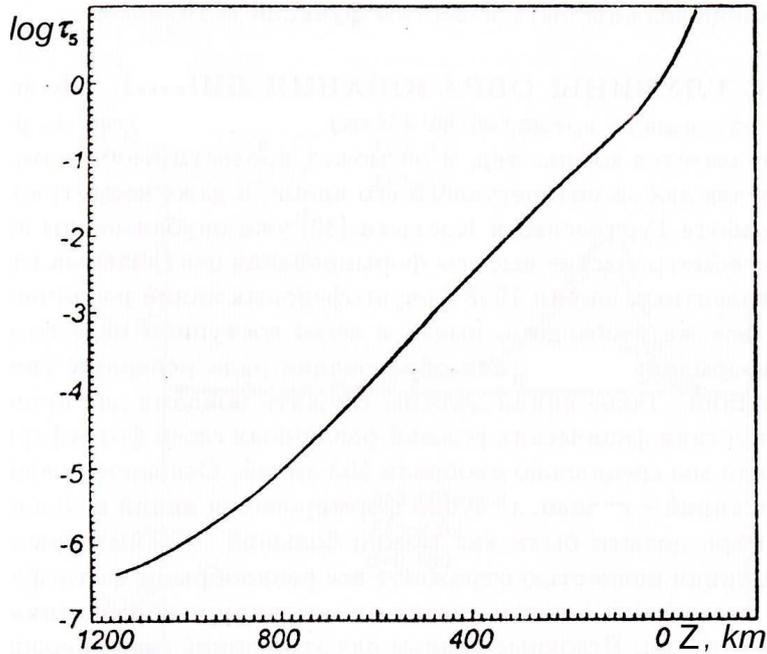}
   \caption
{The optical depth  as a function of the geometrical depth (HOLMU
model)}
  \label {F:5}
  \end{figure}

Our calculations of the CFs and the average depth of formation of
the lines were performed with a software package SPANSAT
\cite{1988ITF...87P....3G}, the solar photosphere model HOLMU
\cite{1974SoPh...39...19H},  the microturbulent velocity $\xi_{\rm
mic}$ = 0.95 km/s, the macroturbulent velocity $\xi_{\rm mac}$  =
1.75 km/s, and the damping constant $1.5\gamma_{\rm wdw} $. The
abundance of chemical elements and line oscillator strengths
($Agf$) were determined in the process of calculation by comparing
the observed and calculated central depths of lines. The line
parameters: wavelength $\lambda $, excitation potential EP, and
observed central line depth $d=(I_c-I_l)/I_c$ were taken from the
tables of Gurtovenko and Kostyk \cite{1989KiIND.........G}. The
results obtained are presented in  a Table. It lists the
parameters of the Fraunhofer lines and the average depths of
formation of  line depression and line emission calculated at the
centre and half-width of line profile as well as  the nearby
continuum. All calculations  are performed  at the centre of solar
disk. The average depths calculated are given in the scale
$\log\tau_5$. To convert the optical depth to the geometrical
depth, we present the dependence of the optical depth $\log\tau_5$
on geometrical depth $z$ in Fig.~\ref{F:5}. The effective width is
given in the geometrical depth.

  \begin{figure}
    \centering
   \includegraphics[width=10cm]{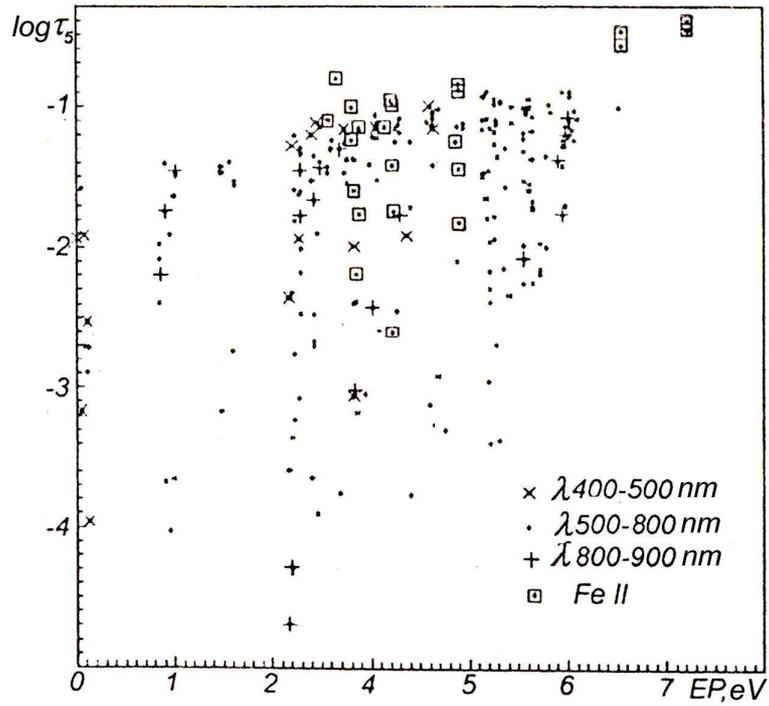}
   \caption
{The average depth of line formation  $\log \tau_5$ vs. the
potential excitation EP for iron lines. The data for Fe II lines
are marked by squares, while  the data for the Fe I line are
marked by the other icons. Large scatter of the values of
formation depths  is caused the difference in the line parameters,
such as the wavelength and central line depth }
  \label {F:6}
  \end{figure}
  \begin{figure}
    \centering
   \includegraphics[width=10cm]{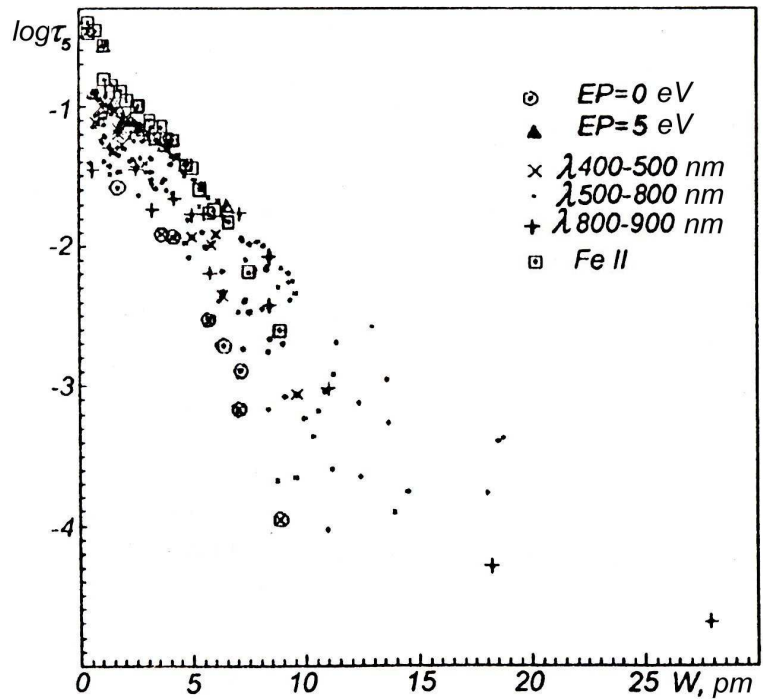}
   \caption
{The average depth of line formation  $\log \tau_5$ vs. equivalent
width $W$}
  \label {F:7}
  \end{figure}
  \begin{figure}
    \centering
   \includegraphics[width=10cm]{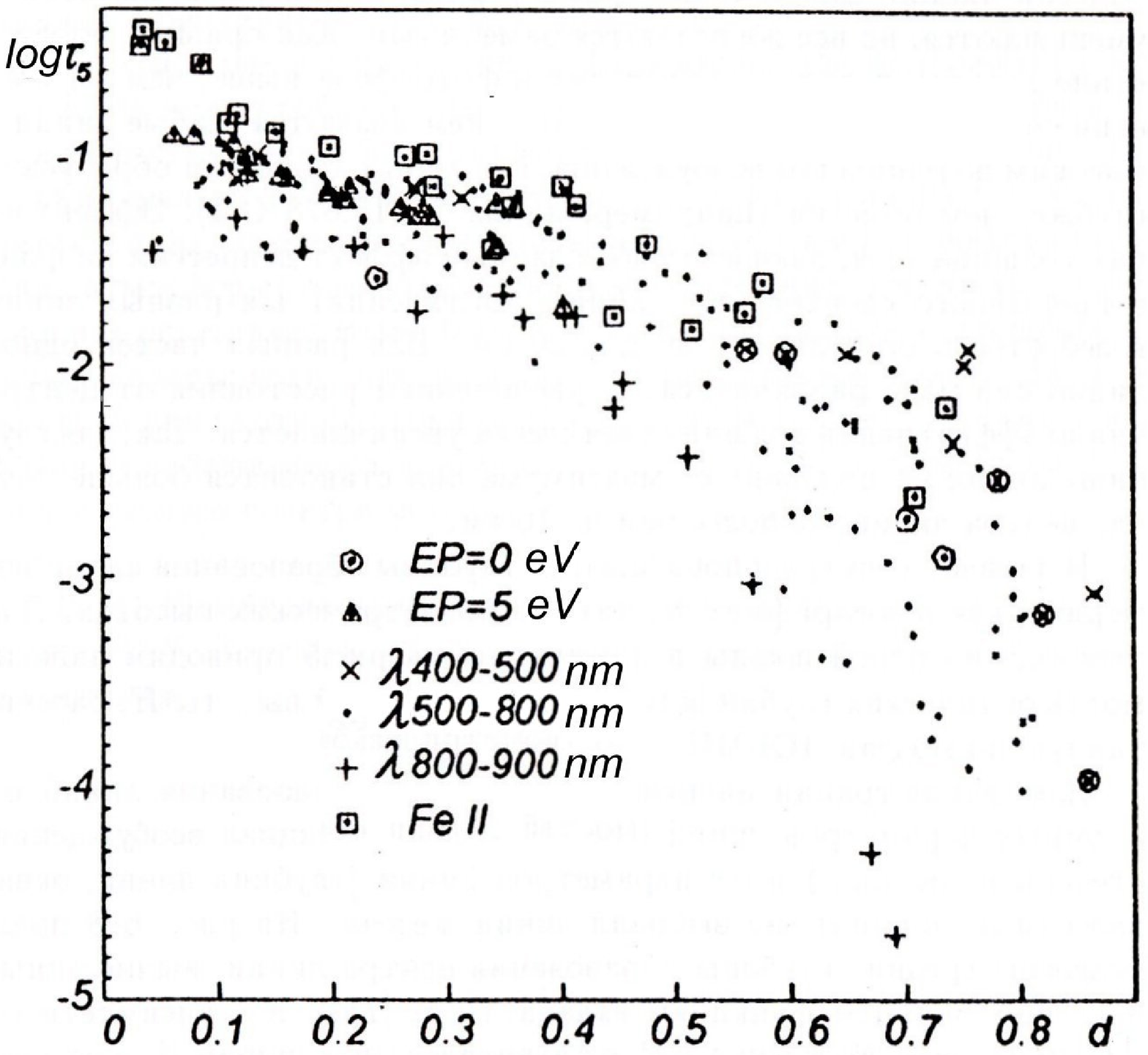}
   \caption
{The average depth of line formation  $\log \tau_5$ vs. central
line depth $d$}
  \label {F:8}
  \end{figure}

On the average, the  depth  of formation of the continuum equals
$\log \tau_5 \approx 0$. It slightly depends on the wavelength due
to the nonlinear wavelength dependence of the absorption
coefficients. The average  depth of formation of the continuum
changes from $\log \tau_5=0.163$  to 0.001  in the wavelength
range from 400.0  to 520.0 nm and from $\log \tau_5=0.001$ to
$-0.215$ in the wavelength range from 520.0 to 900.0 nm.  This
effect is easy to trace in Figs~\ref{F:2}--\ref{F:4}.

The average depth of formation of the very weak line  may be even
greater than the average depth of formation of the nearby
continuum. We found the  597.88 Si II line with EP = 10.07 eV and
${<}\log \tau_5{>}{=}-0.062$, while  the nearby continuum has
${<}\log \tau_5{>}{=}-0.088$. This is not a paradox, but a natural
feature of very weak lines with  high EP.  The far wings of
specific very strong lines are also formed in the deepest layers
of the photosphere (e.g., the H and K  Ca II lines).

In the Table there are few strong lines which are formed in the
region of  temperature minimum. This is the 455.40 Ba II line
(${<}\log \tau_5{>}{=}-5.757$),  the 649.69 Ba II line (${<}\log
\tau_5{>}{=}-4.915$), the 868.86 Fe I line (${<}\log
\tau_5{>}{=}-4.826$), and the 832.70 Fe I line (${<}\log
\tau_5{>}{=}-4.293$). A lot of lines with  ${<}\log
\tau_5{>}{=}-3$ to $-4$ are formed in the upper photosphere, but
the vast majority of lines with ${<}\log \tau_5{>}{=}-0.5$ to $-2$
are formed in  the middle of the photosphere.

Figs~\ref{F:6}--\ref{F:8}  demonstrate the average depth of
formation of the center of  iron lines  depending on the
excitation potential EP, equivalent width $W$, and the
relative depression at the line center $d$ (or central line
depth.) With the help of these Figures it is possible to  estimate
the average depth of iron lines, which are absent in our Table.

Note also that  the  effective widths $\Delta z_{\rm eff}$
calculated for different  Fraunhofer lines  differ from each
other. They vary in the range 50--120 km from one line to another,
while the differences from one point to another  within the line
profile is sufficiently small ($\approx 10$~km).

\section{Conclusion}

We have demonstrated that  line emission and  line depression
contribution functions are different in their essence. The
emission function defines the contributions of elementary
photospheric layers to the emergent continuous  emission and
selective re-emission  at the line frequency, whereas the
depression function defines  the contributions of these layers to
the observed  selective depression and  selective  re-emission  at
the frequency considered. If selectively absorbing  particles are
absent in the photospheric layers, the depression  function is
equal to zero, while the emission function is not zero. The
depression contribution  function characterizes the magnitude of
deficit of the emerging radiation at the line frequency
relatively  the continuum level, i.e, it characterizes the
magnitude of Fraunhofer line. To correctly calculate the average
depth of formation of the Fraunhofer line you should  use  the
depression contribution function derived through the weighting
function.

We examined in detail the physical processes involved in the
formation of the spectral line. The understanding of these
processes  is useful not only for the study of such phenomena as
Fraunhofer lines, but also for the correct selection of lines to
study  of specific events in the solar atmosphere.

     \vspace{0.5cm}
Sheminova V.A.  is sincerely grateful to A.S. Gadun, R.I. Kostyk,
M.Ya. Orlov for useful advice and valuable comments.

 \newpage

  { \normalsize
{\bf Table.} Select spectral lines:   wavelengths ($\lambda$);
excitation potentials of the lower level (EP);   central line
depths ($d$); effective  widths of the line formation region
($\Delta z_{\rm eff}$); average formation depths
(${<}\log\tau_5{>}$) of  emergent line emission $I_l$ and
depression $D_l$ at the line center; emergent line emission and
depression at the line half-width $D_l/2$;   average depth of
formation of the nearby continuum $I_c$. The  average depths were
calculated for the spectral lines at the center of  the solar
disc.      }
{\footnotesize
\begin{longtable}{clrcrrrrrr}
  \multicolumn{10}{c}
{ } \smallskip\smallskip\\
 \hline \\
 $\lambda$&Element& EP& $d$& $\Delta z_{\rm eff}$
 & \multicolumn{5}{c} {${<}\log\tau_5{>}$}\\
 (nm)&&(eV)&& (km)&$I_l$&$D_l$&$I_l/2$&$D_l/2$& $I_c$
 \smallskip\\
   \hline \endfirsthead
 \hline\\
 $\lambda$&Element& EP& $d$& $\Delta z_{\rm eff}$
   & \multicolumn{5}{c} {${<}\log\tau_5{>}$}\\
(nm)&&(eV)&& (km)&$I_l$&$D_l$&$I_l/2$&$D_l/2$& $I_c$  
       \smallskip\\
     \hline
    {continued}  \smallskip\smallskip\\
   \endhead
   670.77  &   Li I    &    0.00 &   .010& 109 &   -0.147&   -1.180 &  -0.146&  -1.132&  -0.144   \\
   313.10  &   Be II   &    0.00 &  .720 & 129 &   -0.193&    -1.238&    0.098&  -0.772&  0.181   \\
   481.73  &   C I     &    7.48 &  .138 &  60 &   -0.139&   -0.253 &  -0.051 &  -0.139&   0.047   \\
   505.21  &   C I     &    7.68 &  .252 &  63 &   -0.361&   -0.411 &  -0.180 &  -0.207&   0.015   \\
   538.03  &   C I     &     7.68&   .158&   62&   -0.274&   -0.337 &  -0.157 &  -0.193&   -0.025  \\
   658.76  &   C I     &     8.53&   .083&   59&   -0.307&   -0.299 &  -0.230 &  -0.179&   -0.136  \\
   711.31  &   C I     &     8.64&  .118 &  60 &   -0.430&   -0.395 &  -0.314 &  -0.236&   -0.169  \\
   833.51  &   C I     &     7.68&   .355&   78&   -1.084&   -1.095 &  -0.638 &  -0.547&   -0.206  \\
\end{longtable}
            }
     \end{document}